\documentclass[sigconf]{acmart} 
\usepackage{pdflscape}
\usepackage{tabularx}
\usepackage{rotating}
\usepackage[indentfirst=false,font={itshape},begintext=``,endtext='']{quoting}
\usepackage{subcaption}
\usepackage{hyperref}
\usepackage{multirow}
\usepackage[most]{tcolorbox}
\usepackage{tabularx}
\newtcolorbox{summarybox}[2][]{
    enhanced,
    breakable,
    sharp corners,
    boxrule=0.3pt,
    colback=white,
    colframe=black!40,
    coltitle=black,
    fonttitle=\bfseries,
    toptitle=2mm,
    bottomtitle=2mm,
    colbacktitle=white,
    title=#2,
}
\usepackage{graphicx}
\usepackage[multiple]{footmisc}
\usepackage{xspace}

\AtBeginDocument{
  \providecommand\BibTeX{{
    \normalfont B\kern-0.5em{\scshape i\kern-0.25em b}\kern-0.8em\TeX}}}

\makeatletter

\usepackage{amssymb}
\newcommand{\ygg@basicalert}[2]{\textcolor{red}{\fbox{\bfseries\sffamily\scriptsize#1}{\sf\small$\blacktriangleright$\textit{#2}$\blacktriangleleft$}}}
\newcommand{\YANN}[1]{\ygg@basicalert{YANN}{#1}}

\begin{document}

\title{Video Game Project Management Anti-patterns}
\author{Gabriel C. Ullmann}
\email{g_cavalh@live.concordia.ca}
\orcid{0000-0002-3274-0789}
\author{Cristiano Politowski}
\email{c_polito@encs.concordia.ca}
\orcid{0000-0002-0206-1056}
\author{Yann-Ga\"el Gu\'{e}h\'{e}neuc}
\email{yann-gael.gueheneuc@concordia.ca}
\orcid{0000-0002-4361-2563}
\affiliation{
  \institution{Concordia University}
  \city{Montreal}
  \state{Quebec}
  \country{Canada}
}

\author{Fabio Petrillo}
\email{fabio@petrillo.com}
\orcid{0000-0002-8355-1494}
\affiliation{
  \institution{Université du Québec à Chicoutimi}
  \city{Chicoutimi}
  \state{Quebec}
  \country{Canada}
}

\author{João Eduardo Montandon}
\email{joao.montandon@dcc.ufmg.br}
\orcid{?}
\affiliation{
  \institution{Universidade Federal de Minas Gerais}
  \city{Belo Horizonte}
  \state{Minas Gerais}
  \country{Brazil}
}

\begin{abstract}
Project Management anti-patterns are well-documented in the software-engineering literature, and studying them allows understanding their impacts on teams and projects. The video game development industry is known for its mismanagement practices, and therefore applying this knowledge would help improve game developers' productivity and well-being. In this paper, we map project management anti-patterns to anti-patterns reported by game developers in the gray literature. We read 440 postmortem problems, identified anti-pattern candidates, and related them with definitions from the software-engineering literature. We discovered that most anti-pattern candidates could be mapped to anti-patterns in the software-engineering literature, except for Feature Creep, Feature Cuts, Working on Multiple Projects, and Absent or Inadequate Tools. We discussed the impact of the unmapped candidates on the development process while also drawing a parallel between video games and traditional software development. Future works include validating the definitions of the candidates via survey with practitioners and also considering development anti-patterns.
\end{abstract}


\newcommand{\vgap}{Video Game Anti-pattern\xspace}
\newcommand{\sap}{Software Anti-pattern\xspace}
\newcommand{\pmap}{Project Management Anti-pattern\xspace}
\newcommand{\dataAPs}{https://?\xspace}
\newcommand{\dataProblems}{https://?\xspace}
\newcommand{\problemsTotal}{418}
\newcommand{\rev}[1]{{\color{red} #1}}
\newcommand{\card}[8]{
    \vspace{0.3cm}
    \noindent {\textbf{#1}} \\[0.05cm]
    \noindent \textsc{Name:} {#2} \\[0.1cm]
    \noindent \textsc{Anti-pattern problem:} {#3} \\[0.03cm]
    \noindent \textsc{Refactored solution:} {#4} \\[0.3cm]
    \noindent \textsc{Case example:} {#5} \\[0.03cm]
    \noindent \textsc{Problem:} {#6} \\[0.03cm]
    \noindent \textsc{Solution:} {#7} \\[0.03cm]
    \noindent \textsc{Full postmortem:} \url{#8}
}

\maketitle

\section{Introduction}

\paragraph{Context:}  A software anti-pattern is a solution to a problem that yields negative consequences \cite{Brown1998}. They exist for \textit{Development}, \textit{Architecture}, and \textit{Project Management}. The latter are related to software processes, human resources, and communication. As defined by \citet{Brown1998}, ``[project] management anti-patterns describe how software projects are impaired by people issues, processes, resources, and external relationships''. They describe scenarios where human communication and relationship issues affect software development processes \cite{Laplante2006}.

\paragraph{Problem:} Over the past years, many project management issues have been reported in the video game development industry\footnote{\url{https://www.bloomberg.com/news/articles/2021-11-17/activision-atvi-ceo-bobby-kotick-is-under-pressure-to-resign}}\footnote{\url{https://www.businessinsider.com/video-game-development-problems-crunch-culture-ea-rockstar-epic-explained-2019-5}}\footnote{\url{https://www.bloomberg.com/news/articles/2022-01-03/bioshock-creator-s-next-game-and-its-narrative-legos-in-turmoil}}. These were not isolated issues: 45\% of problems documented in postmortems are related to management practices \cite{Politowski2021b}. However, while some studies present data related to video game practices \cite{Petrillo2010} and development anti-patterns \cite{Doran2017}, currently there is no discussion about video game project management anti-patterns.

\paragraph{Objective \& Research Questions:} Our main goal is to find out how project management anti-patterns in the video game industry relate to traditional software-engineering anti-patterns. Specifically, we answer the following research question: \emph{how do video game project management anti-patterns fit into the software-engineering anti-patterns literature?}

\paragraph{Method:} We read 440 video game postmortem problems and, based on their description, grouped them into anti-pattern candidates, which are concise anti-pattern proposals. We then mapped these proposals to project management anti-patterns documented in the software-engineering literature. 

\paragraph{Results:} Among the most common project management anti-patterns found in video game development, the most frequent was \textit{Project Mismanagement} \cite{Brown1998}, followed by \textit{Death March} \cite{Yourdon2003}, \textit{Shoeless Children} \cite{Laplante2006}, \textit{Cover Your Assets} \cite{Brown1998}, and \textit{False Surrogate Endpoint} \cite{Cunningham2010}. \textit{Feature Creep} and \textit{Feature Cut} are the most common unmapped anti-pattern candidates, followed by \textit{Absent or Inadequate Tools} and \textit{Working on Multiple Projects}. While some mappings highlight similarities between video game and traditional software development, the absence of others shows the need for studying anti-patterns in the context of video game development.

\paragraph{Conclusion:} Our data shows that some project management anti-patterns happen in both fields: traditional software and video games. In future works, we plan to validate the anti-pattern candidate definitions via a survey to understand if they are accurate, considering a broad range of video games and professionals. We also intend to analyze video game development anti-patterns.

\paragraph{Paper summary:} Section \ref{sec:method} describes the methods of data collection and inference. Section \ref{sec:results} discusses the analysis of the dataset and the relations between variables. Section \ref{sec:discussion} discusses the results. Section \ref{sec:threats} summarises threats to validity. Section \ref{sec:related} describes related works. Section \ref{sec:conclusion} concludes with future work.

\section{Method}
\label{sec:method}

\begin{figure*}[ht]
    \centering
    \includegraphics[width=0.8\linewidth]{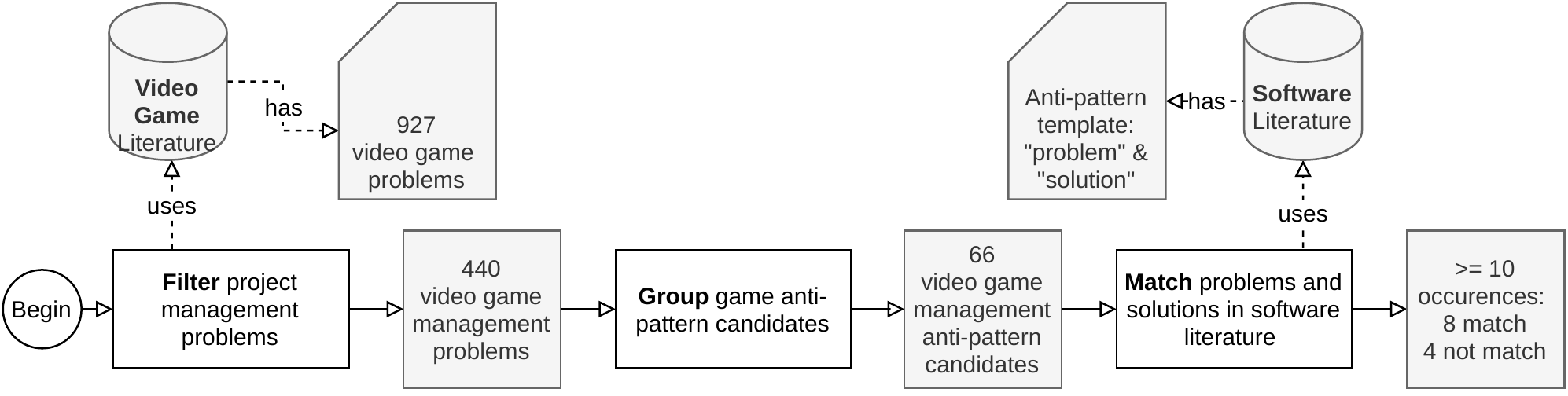}
    \caption{From the video game problems dataset \cite{Politowski2021b}, we filtered out those not related to management. From these, we grouped anti-pattern candidates. After that, we mapped the candidates to the software literature \cite{Brown1998, Laplante2006}.}
    \label{fig:method}
\end{figure*}

We divide our method into three steps, which we perform sequentially, as shown in \autoref{fig:method}.
We describe these steps in the remainder of this section.

\subsection{Dataset of Video Game Problems}
\label{sec:dataset}

We used a dataset of video game problems \cite{Politowski2020a,Politowski2021b}, which contains a curated list of problems gathered from postmortems. It comprises 200 projects from 1997 to 2019, describing 927 problems. Each problem is composed of the video game name, year of release, and a quote, which is a small excerpt from the postmortem describing an issue that occurred during development.

These problems are grouped in four categories: \textit{Production}, \textit{People Management}, \textit{Feature Management}, and \textit{Business}. We considered only \textit{People/Feature Management} problems, which correspond to 45\% of the problems.

\begin{itemize}
    \item \textit{Production} describes operational problems that happen during the production phase;
    \item \textit{People Management} describes management problems related to teams and stakeholders;
    \item \textit{Feature Management} describes management problems related to feature planning, scheduling, and development;
    \item \textit{Business} relates to marketing and strategies to generate revenue. It is not considered a management problem in this analysis.
\end{itemize}

\subsection{Selecting Project Management Problems}

Once we defined the dataset as shown in \autoref{sec:dataset}, we filtered out the problems to include only those related to Project Management. This choice allowed us to reduce our scope while focusing on the category with the most occurrences in the dataset - 440 distinct problems.

\subsection{From Problems to Anti-pattern Candidates}

We read the quotes from the filtered dataset, isolating their descriptions of problems and solutions. For problems in which the quote did not give us enough information, we read the whole postmortem. This process was first done by the main author, and subsequently reviewed iteratively by the others in order to make sure descriptions were concise and well defined for each case.

Then, we applied open card sorting iteratively to group the problems. At first, the main author grouped the problems independently. After finishing this step, all authors discussed together, and sorting changes were made in common agreement. The sorting was considered finished when all authors reached an agreement.

We performed the sorting based on the content of the quotes. We grouped quotes when they shared the same problems and solutions. For example, we found several postmortem quotes mentioning financial problems and how developers reacted to the situation through working harder or reducing expenses, which we grouped into a ``Underbudgeted Project'' anti-pattern candidate. 

Thus, we obtained 66 groups or anti-pattern candidates. We structured these candidates using the mini-template for anti-patterns described by \citet{Brown1998}, who provided three types of anti-pattern templates. These vary according to the amount of information that they provide. For example, the \textit{Pseudo anti-pattern template} shows only the name and the problem while the \textit{Full anti-pattern template} enumerates various attributes. We chose to work with the intermediate \textit{Mini anti-pattern template}, which contains:

\begin{itemize}
    \item \textbf{Name}: What shall this anti-pattern be called by practitioners (pejorative)?
    \item \textbf{Anti-pattern Problem}: What is the recurrent solution that causes negative consequences?
    \item \textbf{Refactored Solution}: How do we avoid, minimize, or refactor the anti-pattern problem?
\end{itemize}

\subsection{From Anti-pattern Candidates to Software Engineering Literature}

For each anti-pattern candidate, we manually searched for a fitting project management anti-pattern in the software-engineering literature \cite{Brown1998, Laplante2006, Brada2019}. For example, we related ``Underbudgeted Project'' to the Shoeless Children anti-pattern from \citet{Brown1998} because they both describe the same problem (deprivation of resources) and solution (strategic budget-cutting that excludes essential infrastructure). Table \ref{tab:mapping-process} show more examples of the mapping.

Not all anti-pattern candidates could be mapped to a software-engineering counterpart. We describe them in Section \ref{sec:results} and discuss their causes and consequences in-depth in Section \ref{sec:discussion}.

\section{Results}
\label{sec:results}

\begin{figure}[ht]
\centering
    \includegraphics[width=.6\linewidth]{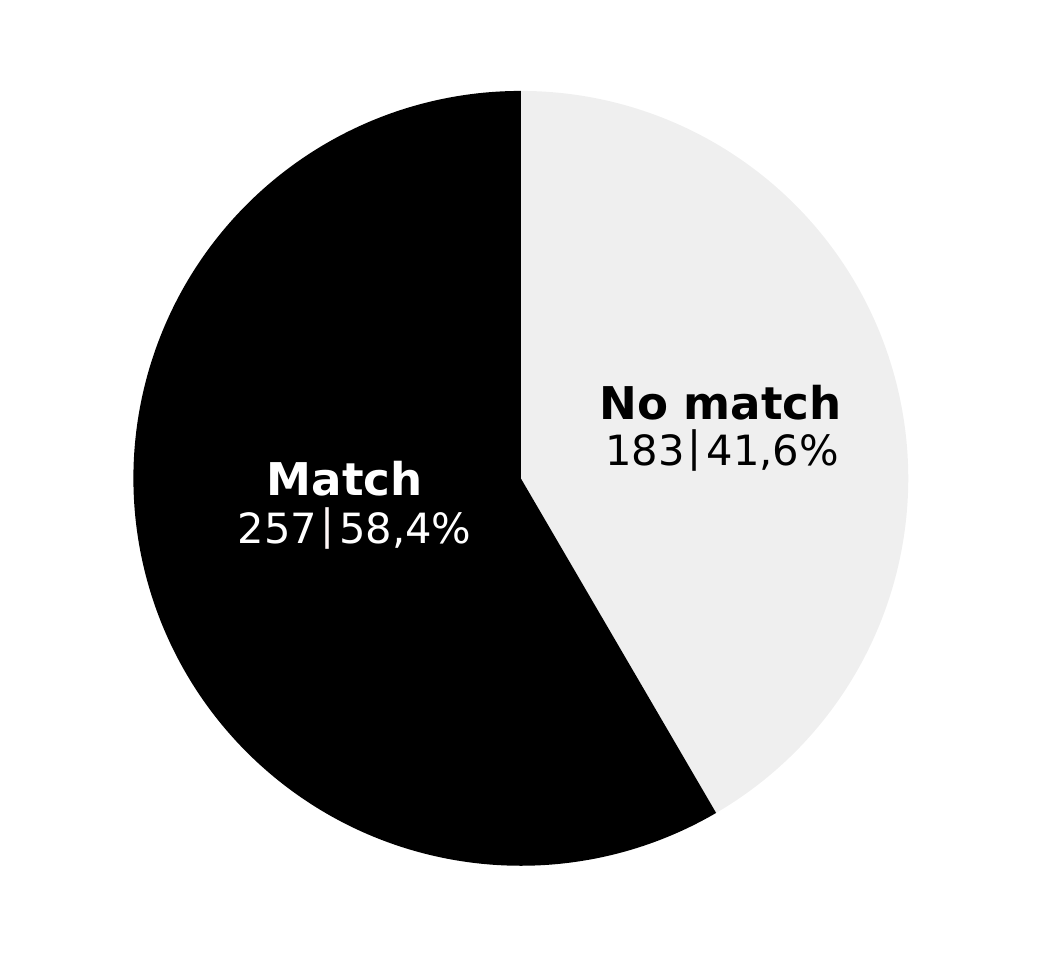}
     \caption{Number of postmortem problems that could be mapped to software literature anti-patterns}
     \label{fig:pie-total}
\end{figure}

From the 440 management problems in the video game literature, we derived 66 different anti-pattern candidates. \autoref{fig:pie-total} shows that we could map 57\% of the candidates to software-engineering anti-patterns.

Among the 66 anti-pattern candidates, we found 12 with a number of occurrences equal to or greater than 10 (\autoref{tab:common-aps}). Four candidates were not in the software-engineering literature: \textit{Feature Creep}, \textit{Feature Cuts}, \textit{Working on Multiple Projects}, and \textit{Absent or Inadequate Tools}. The next section describes these anti-pattern candidates. Due to the lack of space, the candidates that could be mapped into the software-engineering literature are available on-line\footnote{\url{https://zenodo.org/record/5828315}}.

\begin{table*}[ht]
\centering
\small
\caption{The most common anti-patterns in video game literature (postmortems) and its mapped software anti-patterns.}
\label{tab:common-aps}
\begin{tabular}{@{}rr|lll@{}}
\toprule
Video game AP candidate & Occ. & Software AP & AP Category & Source \\ \midrule
Underestimated Tasks & 37 & Project Mismanagement & Management & \citet{Brown1998} \\
Feature Creep & 33 & - & - & - \\
Death March & 30 & Death March & Management & \citet{Yourdon2003} \\
Feature Cuts & 27 & - & - & - \\
Understaffed Team & 26 & Project Mismanagement & Management & \citet{Brown1998} \\
Underbudgeted Project & 23 & Shoeless Children & Environmental & \citet{Laplante2006} \\
Neglected Game Design & 13 & Cover your Assets & Architecture & \citet{Brown1998} \\
Working on Multiple Projects & 13 & - & - & - \\
Absent or Inadequate Tools & 12 & - & - & - \\
Testing/QA Issues & 11 & Project Mismanagement & Management & \citet{Brown1998} \\
False Surrogate Endpoint & 10 & False Surrogate Endpoint & Management & \citet{Cunningham2010} \\ 
Lack of Technical Knowledge/Planning & 10 & Project Mismanagement & Management & \citet{Brown1998} \\ \bottomrule
\end{tabular}
\end{table*}

\begin{table*}[ht]
\centering
\small
\caption{Example of the mapping process for problems, anti-pattern candidates and the literature}
\label{tab:mapping-process}
\begin{tabular}{@{}
    p{\dimexpr.05\textwidth}
    p{\dimexpr.27\textwidth}
    p{\dimexpr.10\textwidth}
    p{\dimexpr.10\textwidth}
    p{\dimexpr.27\textwidth}@{}}
\toprule
 & \textbf{Postmortem description} & \textbf{Anti-pattern candidate} & \textbf{Relates to} & \textbf{Literature description} \\ \midrule
Problem & \(\cdot\) ``Golden Abyss had 34 next-gen quality levels, each of which had five times the content of any PSP game we had ever made'' \newline \(\cdot\) ``From January to December of 2011, the team worked pretty much nonstop (...)'' & Death March
 & Death March (Yourdon, 2003)
 & \(\cdot\) The schedule has been compressed to less than half the amount estimated by a rational estimating process \newline \(\cdot\) The functionality or other technical aspects of the project are twice what they would be under normal circumstances \\ [2cm]
Solution & \(\cdot\) ``Negotiating as reasonable a schedule as possible (...)'' \newline \(\cdot\) ``Holding weekly meetings with every individual on the team, where issues (...) can be raised directly.'' \newline \(\cdot\) ``Vetting the scope of our next project up front (...)'' &   &   & \(\cdot\) Agreement on interfaces \newline \(\cdot\) Peer reviews \newline \(\cdot\) Metric-based scheduling and management \newline \(\cdot\) Formal risk management \\
\midrule
Problem & \(\cdot\) ``Because we didn\'t have a firm design, it was impossible to do accurate time estimates'' \newline \(\cdot\) ``The only way we could get the game done on time was to cut deeply into our testing schedule''
& Underestimated Tasks 
 & Project Mismanagement (Brown et al, 1998)
 & ``Often, key activities are overlooked or minimized. These
include technical planning (architecture) and quality--control activities (inspection and test)''\newline \\ 
 [2cm]
Solution & ``Our current focus is on getting solid designs done up front and solid testing done on the back end'' &   &   & \(\cdot\) Prototyping \newline \(\cdot\) Create and document a risk management analysis \newline \(\cdot\) Define architectural dependencies \\
\midrule
Problem & ``(...) the original game lacked a truly focused design. We knew what the fundamentals of the game would be, but we did not have the specifics that we needed to create a solid, cohesive product.''
& Neglected Game Design 
 & Cover your Assets (Brown et al, 1998)
 & ``Document--driven software processes often employ authors who list alternatives instead of making decisions'' \newline \\[2cm]
Solution &  \(\cdot\) ``We created total level walkthroughs, with each room and encounter written out'' \newline \(\cdot\) `` Flowcharts were used to draw the preliminary levels, and concept art was used for key location elements'' &   &   & Creating architectural blueprints, which are ``a small set of diagrams and tables that communicate the operational, technical, and systems architecture of current and future information systems.'' \\
\bottomrule
\end{tabular}
\end{table*}

\card{Feature Creep (33 cases)}{``Feature Creep''.}{Features keep being added to the project, expanding its initial scope excessively. While the new features might be worthy additions to the game, they might also overwhelm the game developers and distract from core features.}{Aiming for clear game design requirements and definitions from the beginning. Even though the development is iterative and features may change, keep them in check to avoid losing control.}{Baldur's Gate 2, 2001}{``While we shipped with nearly all the features we originally planned, we did start cutting quests and characters well before the final testing phase. We still ended up with over 200 hundred hours of gameplay.''}{``In retrospect we should have started this process many months earlier. One of the dangers of development is that game developers have a tendency to always add content if they are given time. (...) Its wise to use that prioritized feature list to hone the work (of course ours was informal, which made it a little difficult)''}{https://www.gamedeveloper.com/design/-i-baldur-s-gate-ii-i-the-anatomy-of-a-sequel}

\card{Feature Cuts (27 cases)}{``Death by a Thousand Cuts'', ``Ripple Effect'', ``Domino Effect''}{A video game project enters production with a well-defined feature set, but without proper resource allocation and priority definitions. Features may be partially developed, but when more in-depth planning is made, the management decides to cut some of them to save time and--or money.}{Prototyping and playtesting can help designers and developers understand what features are valuable to players and thus prioritise them. When cuts are unavoidable, discussing and documenting them with the participation of all developers is important to avoid excessive re-work.}{The Political Machine, 2004}{``We had several other features in mind that we had to cut because of lack of time, time we would have had if we hadn't done multiplayer. (...) When gamers demand multiplayer, what they don't realize is that something has to be sacrificed for it. You have a finite budget and a finite time to use it (...)''}{``I think the inclusion of different maps to play on would have increased the replayability of the game a great deal. (...). But the time in development and testing on multiplayer eliminated those kinds of things. (...) With that in mind, I wish we hadn't done multiplayer in The Political Machine.''}{https://www.gamedeveloper.com/disciplines/postmortem-stardock-s-i-the-political-machine-i-}

\card{Working on Multiple Projects (13 cases)}{``Divide and Not Conquer'', ``False Parallelism''}{Game developers must work multiple jobs, often for financial reasons, or manage several activities inside their company, taking a workload much higher than normal. They become overwhelmed and misaligned with project goals over time.}{Analysing the time availability of each game developer on the team and carefully managing resource allocation may be a way to mitigate this issue. Staffing up may also be an option when budget and structure allow.}{Rollers of the Realm, 2014}{``Having the production timeline stretch out meant Rollers rubbed against the work-for-hire projects that we lined up because we need to keep the studio lights on. (...)''}{``We needed to split our resources into two project teams. Phantom Compass often works on multiple projects at once, and in many ways this is a strength, but in this case it caused more strain on finishing up Rollers than we would have liked (...).''}{https://gamedeveloper.com/design/postmortem-pinball-rpg-hybrid-i-rollers-of-the-realm-i-}

\card{Absent or Inadequate Tools (12 cases)}{``Bringing a Knife to a Gun Fight'', ``Wrong Tool for the Job''}{Developers and--or artists begin production without having a standardized set of tools for their activities or using outdated/improvised tools. They do so due to management inexperience, lack of time, budget, or qualified staff.}{Looking for third-party tools that fit the project's needs or developing custom in-house tools are activities that must be carefully considered early in the project.}{Swords \& Soldiers, 2010}{``The levels took a lot of time to build with the limited tools we had. With only one programmer, you cant ask for a good level design tool For design, then, we had to use the stock Windows text editor, Notepad.''}{``Our AI routines were built in a simple logic tree editor, which was built by a programming intern. And though the editor was a bit buggy, it was still a major improvement over editing those XML files in Notepad. (...) In hindsight, we probably should have put some time in coding creating a visual level editor.''}{https://www.gamedeveloper.com/audio/postmortem-ronimo-games-i-swords-soldiers-i-}

\section{Discussion}
\label{sec:discussion}

\subsection{Mapped Anti-pattern candidates}

Project Mismanagement, Underestimated Tasks and Death March are the top-3 most frequent. Most of the anti-pattern candidates were mapped to the Project Mismanagement anti-pattern, which describes a broad range of situations. 

Knowing the prevalence of these anti-patterns can help project managers and product owners proactively monitor their development teams and correct their actions accordingly.

The mapping also shows the difficulty faced by game developers in creating a well-organized and stable work environment. Such issue has been frequently addressed in the literature \cite{cote2021,edholm2017,Petrillo2009}. Our results reinforce this issue.

\subsection{Unmapped Anti-patterns candidates}

The most important contribution of this work resides in the anti-pattern candidates not mapped into the software-engineering literature. Researchers should conduct further studies to better understand in which scenarios they arise, as well as solutions to solve or mitigate them. Researchers could also rely on this data to build a video game-oriented anti-pattern catalog.

We now discuss the circumstances in which these anti-patterns happen, contrasting the experiences of game developers and traditional software developers.

\subsubsection{Feature Creep}

In the context of traditional software, Elliot \cite{Elliott2007} argues that Feature Creep ``is a natural result of building something new and exploring new territory''. We argue this is also true for video game development, especially when looking for the fun factor, as described by the creators of \textit{Matt Hazard: Blood Bath and Beyond}:``it took longer to feel out the fun factor of our game than we had anticipated, we crept over our allotted budget on both time and money''. \textit{Command and Conquer} creators also stated that ``a team has to be able to incorporate new ideas during development'' but must also ``be able to cut features diplomatically when it is in the best interest of the project''.

Additionally, Feature Creep and Feature Cuts are reported by Petrillo et al. \cite{Petrillo2009} to be the problems most commonly described in postmortems. Politowski et al. \cite{Politowski2021b} report that "feature creep, and, to a less degree, delays, are recurring problems during game development". Therefore, even though this anti-pattern is not referenced by \citet{Brada2019} in their catalog, we argue it is not a game-specific issue, since it is widely known both in video games and traditional software industries.

\subsubsection{Feature Cuts}

Similarly, the ``Feature Cut" anti-pattern candidate is well-known. It distinguishes from ``Feature Creep" because it describes the removal of features that were included in the initial scope, and not included afterward. When ``Feature Creep" occurs in a video game project, the team may choose not to cut down the newly-added features and just move on with the extra workload. \textit{Fallout Tactics} creators describe that ``rather than reduce the size of the maps or lose maps altogether, we decided to press on and fill them with interesting objectives". However, the choice of adding/removing features is a common challenge in any project so we also argue it is not a game-specific problem.

\subsubsection{Absent or Inadequate Tools}

The recurrence of the ``Absent or Inadequate Tools" anti-pattern candidate reflects the impact of tools in video game production. For example, \textit{Brutal Legend} manager Caroline Esmurdoc states that "good tools needed to be a priority to ensure faster iteration times for our team". \textit{Swords \& Soldiers} creators say that developing a better AI scripting tool "would have saved the rest of the team a lot of time". 

Different from traditional software, video game production encompasses several art processes: animation, graphics, and sound edition. In the postmortems, these tools are pointed out as the most frequent source of problems when they are not properly chosen and standardized.

Tool maturity also has an impact on project stability. Game engines, for instance, are continuously changing their internal APIs and systems to adapt to new GPU features, gaming hardware, and operating systems. Such upgrades may cause bugs or require developers to refactor code written for previous versions of the tool.

Additionally, game engines are not as varied as traditional software IDEs. Unreal and Unity are used by the wide majority of commercial projects \cite{Politowski2021} and this has an influence on professionals' learning choices. Since education institutions focus on courses for these popular tools, the developer's knowledge may not be broad enough, which makes it harder for them to choose or develop their own tools. Therefore, we argue this candidate is tightly related to video games.

\subsubsection{Working on Multiple Projects}

Especially in the indie scene, we can observe that the ``Working on Multiple Projects" anti-pattern candidate is very recurrent since the informality and time flexibility of the production process allows it. For example, while one of the Mini Metro programmers worked full-time as a banking software developer, ``Sportsfriends" developer Douglas Wilson says that he was ``finishing up his PhD" concurrently with the development of the game. However, both video game studios and software companies may work on multiple projects simultaneously, and thus we cannot argue this is a game-specific anti-pattern candidate.

\section{Threats to Validity}
\label{sec:threats}

\paragraph{Limited coverage of problem databases:} The database of postmortems on which our research was based does not represent the whole industry or all issues that may arise in the management of a video game development team. Additionally, the anti-pattern catalog may also not represent the set of all documented issues of this kind, and in some cases, it does not provide accurate sources to deepen research. 
\paragraph{Not all pairs of problems/solutions are documented:} Since postmortems are informal memoirs of a video game's development, they do not always provide all details on how a problem has manifested and how it has been mitigated or solved. Consequentially, we cannot state which problems represent a fully-fledged anti-pattern. Similarly, there is no documented problem/solution description for all software anti-patterns. Especially on those documented by Cunningham \cite{Cunningham2010}, there is often only a problem description and little to no discussion regarding solutions.
\paragraph{Problems and anti-patterns were manually mapped:} We relied on our interpretation of the problem quotes to group them into anti-pattern candidates, and also to relate these candidates to software anti-patterns. Aiming to reduce bias, we read the postmortems in full when these quotes did not provide enough context on both problem and solution. We then made several iterations of discussion, updating our candidate anti-pattern categorization when necessary.

\section{Related Works} \label{sec:related}

\citet{Brown1998,Laplante2006} were our main reference about software anti-patterns. In their books, they illustrate the overall concepts and provide practical examples. The issue is that some of the descriptions are either too broad, short, or informal to convey a proper definition. 

\citet{Brada2019} shows a more complete and up-to-date review of the software project management anti-patterns, which they refer to as ``process anti-patterns''. They created a open catalog of anti-patterns using various sources\footnote{\url{https://github.com/ReliSA/Software-process-antipatterns-catalogue}}. However, after further investigation, the list is not complete to this day. 

To the best of our knowledge, video game project management anti-patterns are not present in the literature. At least not in the same format of traditional software anti-pattern. 

As for development anti-patterns (related to code), \citet{Borrelli2020} performed a similar work. The authors identified, from the video game literature (textbooks, interviews, gray literature), six types of bad smells related to performance, maintainability, and incorrect behavior problems. After that, they used a static analysis tool on 100 open source video games made with the Unity engine. They found the following bad smells:
\begin{itemize}
    \item Allocating and destroying GameObjects in updates;
    \item Getting a GameObject reference finding it by name;
    \item Heavyweight Update methods;
    \item Lack of separation of concerns;
    \item Coupling objects through the IDE Inspector;
    \item Animation speed depends on the frame rate.
\end{itemize}

\section{Conclusion} \label{sec:conclusion}

In this work, we focused on understanding the relation between anti-patterns that occur in video games and the traditional software industry. Taking postmortem problems as a starting point, we identified anti-pattern candidates, which could be mapped to documented software anti-patterns in the majority of cases. 

Regarding the four anti-pattern candidates that could not be mapped, we concluded that only ``Absent or Inadequate Tools''  closely relate to the game development industry, while the others may happen on any software project. Therefore, our findings show that management problems in video game projects are not exclusive since they also happen in the traditional software industry. 

In future works, we intend to validate the candidate anti-pattern mini-template definitions via survey with video game developers and managers. We will also expand the analysis to include software development anti-patterns (code smells).

\begin{acks}
The authors were partly supported by the NSERC Discovery Grant and Canada Research Chairs programs.
\end{acks}

\bibliographystyle{ACM-Reference-Format}
\bibliography{main.bib}

\end{document}